\documentstyle[12pt]{article}
\begin{document}

\title{Generation of entangled states of two traveling modes for fixed number of photon}
\author{XuBo Zou, K. Pahlke and W. Mathis  \\
\\Institute TET, University of Hannover,\\
Appelstr. 9A, 30167 Hannover, Germany }
\date{\today}
 \maketitle

\begin{abstract}
We present an scheme to generate entangled state between two
traveling modes for fixed number of photon, which is based on beam
splitter transformation and conditional zero photon counters.

\vspace{0.25cm} PACS number(s): 42.50.Dv, 03.67.-a
\end{abstract}
%\pacs{PACS number(s): 42.50.Dv, 03.67.Lx}
%}
It has long realized that the striking nature of entanglement lies
at the heart of study of the fundamental issues in quantum
mechanics as witnessed by Einstein-Podolsky-Rosen paper \cite{ab},
Bell' inequality \cite{be} and its subsequent experimental
verification \cite{fc,as}. The recent surge of interest and
progress in quantum information theory allows one to take a more
positive view of entanglement and to regard it as an essential
resource for many ingenious applications such as quantum
teleportation \cite{bb} and quantum cryptography \cite{ak}. These
applications rely on the ability to engineer and manipulate
entangled states in a controlled way. So far, the generation and
manipulation of entangled states have been demonstrated with ions
in a ion trap \cite{mm} and with atoms in Cavity QED \cite{hm}.
Recently, quantum engineering of light received extensively
attention, since phenomena such as teleportation \cite{bp} and
quantum dense coding \cite{mw} found their implementation in the
quantum domain. Different schemes have been proposed to generate
any superposition of vacuum and one photon state \cite{ap,pp}. A
conditional scheme based on beam splitters and conditional zero
counters has also been suggested to produce any quantum
superposition of traveling waves \cite{dc}. More recently, it has
been shown that single photon entangled states can be used to
implement teleportation of superposition states of zero and one
photon states \cite{hj,vd} and to test quantum nonlocality
\cite{hj}. The scheme for teleportation of traveling wave states,
which is a coherent superposition of N Fock state
\begin{eqnarray}
\Psi=\sum_{n=0}^NC_n|n>\, , \label{(1)}
\end{eqnarray} is proposed
\cite{kk}. As a first step of teleportation, we need to prepare
entangled states
\begin{eqnarray}
\Psi=\frac{1}{\sqrt{N+1}}\sum_{n=0}^N|n>_1|N-n>_2 \,,\label{(2)}
\end{eqnarray}
which are entangled states of two traveling modes for fixed number
$N$ of photons. Such state are also of potential interest in the
context of phase sensitivity in two mode interferometer, where
they should allow measurements at the Heiserberg uncertainty limit
\cite{djw}. Recently it is also shown that such states allow the
sub-diffraction limited lithography \cite{pk}.\\In this paper, we
propose a generation of Ref \cite{dc} procedure for generation of
an arbitray two modes entangled state for fixed number of photons,
which can be expressed as
\begin{eqnarray}
\Psi=\sum_{n=0}^NC_n|n>_1|N-n>_2\, . \label{(3)}
\end{eqnarray}
Using the definition of Fock state, Eq.(\ref{(3)}) can be
expressed as
\begin{eqnarray}
\Psi=\sum_{n=0}^NC_n\frac{a^{\dagger{n}}b^{\dagger{N-n}}}{\sqrt{n!(N-n)!}}|0>\,.
\label{(4)}
\end{eqnarray} Here $|0>$ is vacuum state of two
traveling modes. Similar to Ref \cite{dc}, we rewrite the
Eq.(\ref{(3)}) as
\begin{eqnarray}
\Psi=(a^{\dagger}-\beta_Nb^{\dagger})(a^{\dagger}-\beta_{N-1}b^{\dagger})\cdots(a^{\dagger}-\beta_1b^{\dagger})|0>
\,.\label{(5)}
\end{eqnarray} Here $\beta_1\cdots\beta_N$ are the N
(complex) roots of the characteristic polynomial
\begin{eqnarray}\sum_{n=0}^NC_n\frac{\beta^{{n}}}{\sqrt{n!(N-n)!}}=0\, .
\label{(6)}
\end{eqnarray} The relations
\begin{eqnarray}
a^{\dagger}-\beta{b^{\dagger}}&=&\sqrt{1+|\beta|^2}B(\theta,\varphi)a^{\dagger}B^{\dagger}(\theta,\varphi)
\,;\nonumber\\     %\label{(7)}\\
B(\theta,\varphi)&=&\exp(\theta e^{-i\varphi}a^{\dagger}b-\theta
e^{i\varphi}ab^{\dagger})\,;\nonumber\\
\cos\theta &=&\frac{1}{\sqrt{1+|\beta|^2}}\,;\nonumber\\
e^{i\varphi}&=&\frac{\beta}{|\beta|}\,;\nonumber
\end{eqnarray} are used. The quantum state
\begin{eqnarray}
\Psi&=&\left(\prod_{i=1}^N\sqrt{1+|\beta_i|^2}\right)B(\theta_N,\varphi_N)a^{\dagger}B^{\dagger}(\theta_N,\varphi_N)
\cdots\nonumber\\
&&\qquad\times
B(\theta_2,\varphi_2)a^{\dagger}B^{\dagger}(\theta_2,\varphi_2)
B(\theta_{1},\varphi_{1})\quad a^{\dagger}|0> \label{(8)}
\end{eqnarray}
is obtained. Hence, any entangled quantum state of the form
(\ref{(3)}) can be generated from the one-photon state by this
method, which bases on a succession of alternate single photon
addition and beam splitter transformations. These are determined
by the roots of the
characteristic polynomial (\ref{(6)}).\\
In what follows, we give a scheme to generate quantum states of
the form (\ref{(8)}), based on the conditional zero photon
counters. In Ref \cite{dk} it was shown that a mode prepared in an
arbitrary state $|\Phi>$ is mixed at the beam splitter with a
single photon Fock state. A zero photon measurement is performed
in one of the output channels of the beam splitter. Then the
quantum state of the mode in the other output channel collapses to
$Y|\Phi>$ with $Y=Ra^{\dagger}T^{n_a}$. R is the reflectance and T
is the transmittance of the beam splitter. The implementation of
the present scheme is outlined in Fig.1. Following Ref \cite{dc},
the quantum state, which is generated if no photon is detected in
each of the $N-1$ conditional output measurement, is given by
\begin{eqnarray}
\Psi&\sim&
B(\theta^{\prime}_N,\varphi^{\prime}_N)a^{\dagger}T^{n_a}B(\theta^{\prime}_{N-1},\varphi^{\prime}_{N-1})
a^{\dagger}T^{n_a}\cdots\nonumber\\
 &&\times B(\theta^{\prime}_3,\varphi^{\prime}_3)
a^{\dagger}T^{n_a}B(\theta^{\prime}_2,\varphi^{\prime}_2)a^{\dagger}T^{n_a}
B(\theta^{\prime}_{1},\varphi^{\prime}_{1})a^{\dagger}|0>
\label{(9)}\,.
\end{eqnarray}
In order to bring Eq.(\ref{(9)}) into the form of Eq.(\ref{(8)}),
we write Eq.(\ref{(9)}) as
\begin{eqnarray}
\Psi&\sim& B_Na^{\dagger}B^{\dagger}_NB_NT^{n_a}B_{N-1}
a^{\dagger}B^{\dagger}_{N-1}T^{-n_a}B^{\dagger}_{N}\nonumber\\
&&\times
B_NT^{n_a}B_{N-1}T^{n_a}B_{N-2}a^{\dagger}B^{\dagger}_{N-2}T^{-n_a}B^{\dagger}_{N-1}T^{-n_a}B^{\dagger}_{N}
\cdots\nonumber\\
 &&\times B_NT^{n_a}B_{N-1}T^{n_a}\cdots
B_{2}T^{n_a}B_{1}a^{\dagger}B^{\dagger}_{1}T^{-n_a}B^{\dagger}_{2}\cdots
T^{-n_a}B^{\dagger}_{N-1}T^{-n_a} B^{\dagger}_N|0> \,.\label{(10)}
\end{eqnarray}
The abbreviation $B_i=B(\theta^{\prime}_{i},\varphi^{\prime}_{i})$
is used. By means of the Hausdorff formula, we obtain
\begin{eqnarray}
B_NT^{n_a}B_{N-1}T^{n_a}\cdots
T^{n_a}B_{N-i}a^{\dagger}B^{\dagger}_{N-i}T^{-n_a}\cdots
T^{-n_a}B^{\dagger}_{N-1}T^{-n_a} B^{\dagger}_N
=A_ia^{\dagger}-B_ib^{\dagger}\nonumber\\
B_NT^{n_a}B_{N-1}T^{n_a}\cdots
T^{n_a}B_{N-i}b^{\dagger}B^{\dagger}_{N-i}T^{-n_a}\cdots
T^{-n_a}B^{\dagger}_{N-1}T^{-n_a} B^{\dagger}_N
=C_ia^{\dagger}-D_ib^{\dagger} \,.\label{(11)}
\end{eqnarray}
The abbreviations $A_j$,$B_j$, $C_j$ and $D_j$ are
\begin{eqnarray}
A_{j+1}&=&(T\cos\theta^{\prime}_{N-j-1}A_j-\sin\theta^{\prime}_{N-j-1}e^{i\varphi^{\prime}_{N-j-1}}C_j\,;\nonumber\\
B_{j+1}&=&(T\cos\theta^{\prime}_{N-j-1}B_j-\sin\theta^{\prime}_{N-j-1}e^{i\varphi^{\prime}_{N-j-1}}D_j\,;\nonumber\\
C_{j+1}&=&(\cos\theta^{\prime}_{N-j-1}C_j+T\sin\theta^{\prime}_{N-j-1}e^{-i\varphi^{\prime}_{N-j-1}}A_j\,;\nonumber\\
D_{j+1}&=&(\cos\theta^{\prime}_{N-j-1}D_j-T\sin\theta^{\prime}_{N-j-1}e^{-i\varphi^{\prime}_{N-j-1}}B_j\,;\nonumber\\
%\label{(12)}
A_{0}&=&\cos\theta^{\prime}_{N}\,;\nonumber\\
B_{0}&=&\sin\theta^{\prime}_{N}e^{i\varphi^{\prime}_{N}}\,;\nonumber\\
C_{0}&=&\sin\theta^{\prime}_{N}e^{-i\varphi^{\prime}_{N}}\,;\nonumber\\
D_{0}&=&\cos\theta^{\prime}_{N} \,.\nonumber
\end{eqnarray}
Thus, we obtain
\begin{eqnarray}
\Psi\sim
B(\alpha_N,\beta_N)a^{\dagger}B^{\dagger}(\alpha_N,\beta_N) \cdots
B(\alpha_2,\beta_2)a^{\dagger}B^{\dagger}(\alpha_2,\beta_2)
B(\alpha_{1},\beta_{1})a^{\dagger}\,\,|0> \label{(14)}
\end{eqnarray}
The abbreviations are the following:
\begin{eqnarray}
\alpha_N&=&\theta^{\prime}_{N}\,;\nonumber\\
\beta_N&=&\varphi^{\prime}_{N}\,;\nonumber\\
\cos\alpha_{N-i}&=&\frac{|A_i|}{\sqrt{|A_i|^21+|B_i|^2}}\,;\nonumber\\
e^{i\beta_{N-i}}&=&\frac{B_i|A_i|}{|B_i|A_i}\,.\nonumber
\end{eqnarray}
Comparing Eq.(\ref{(14)}) and Eq.(\ref{(8)}), we find that these
two equations become identical, if the parameters are chosen to
\begin{eqnarray}
\alpha_j=\theta_{j}\,;\qquad \beta_j=\varphi_{j}\,.\label{(16)}
\end{eqnarray} The solution of the Eq.(\ref{(16)}) always exits and we can
indeed prepare
any desired entangled state of form Eq.(\ref{(3)}).\\
In the following, we calculate the probability $P$ of generating a
desired state (\ref{(3)}). Obviously this probability is
determined by the requirement that all the $N-1$ detectors do not
register photons. It can be given by
\begin{eqnarray}
P=P(N-1,0|1,0;2,0;\cdots;N-2,0)\cdots P(2,0|1,0)P(1,0)\,.
\label{(17)}
\end{eqnarray}
Here $P(k,0|1,0;2,0;\cdots;k-1,0)$ is the probability that the
k-th detector does not register photons under the condition, that
the detectors $D_1, \cdots, D_{k-1}$ have also not registered
photons. Starting from
\begin{eqnarray}
P(1,0)=||YB_1|1>||^2 \label{(18)}\\
%\mbox{with}\\
Y=Ra^{\dagger}T^{n_a} \end{eqnarray}. Here
$|||\psi>||=\sqrt{<\psi||\psi>}$. we derive
\begin{eqnarray}
P(k,0|1,0;2,0;\cdots;k-1,0)=\frac{||YB_{k}YB_{k-1}\cdots
YB_{1}|1>||^2}{||YB_{k-1}\cdots YB_{1}|1>||\cdots ||YB_1|0>||}
\,.\label{(19)} \end{eqnarray} Combining Eq.(\ref{(17)}),
Eq.(\ref{(18)}) and Eq.(\ref{(19)}), we obtain
\begin{eqnarray}
P&=&P_{N-1}^2\prod_{k=1}^{N-2}(P_k)^{k+3-N} \,;\label{(20)}\\
%\mbox{with}\\
 P_k&=&||YB_{k}YB_{k-1}\cdots YB_{1}|1>||\,;\\
\left(\frac{P_k}{|R|^k}\right)^{2}&=&\sum_{j_1+\cdots+j_{k}=0}^{k}|(1+j_1+\cdots+j_{k})!(k-j_1-\cdots-j_{k})!\nonumber\\
&&\qquad \qquad\times A_{k+1,1}^{j_1}\cdots
A_{k+1,k}^{j_{k}}B_{k+1,1}^{1-j_{1}}\cdots B_{k+1,k}^{1-j_{k}}
|^2\,. \label{(21)}
\end{eqnarray}
Where $j_i=\in\{0;1\}$ and $A_{k,j}$ and $B_{k,j}$ are determined
by:
\begin{eqnarray}
A_{k,j+1}&=&(T\cos\theta^{\prime}_{k-j-1}A_{k,j}-\sin\theta^{\prime}_{k-j-1}e^{i\varphi^{\prime}_{k-j-1}}C_{k,j}\,;\nonumber\\
B_{k,j+1}&=&(T\cos\theta^{\prime}_{k-j-1}B_{k,j}-\sin\theta^{\prime}_{k-j-1}e^{i\varphi^{\prime}_{k-j-1}}D_{k,j}\,;\nonumber\\
C_{k,j+1}&=&(\cos\theta^{\prime}_{k-j-1}C_{k,j}+T\sin\theta^{\prime}_{k-j-1}e^{-i\varphi^{\prime}_{k-j-1}}A_{k,j}\,;\nonumber\\
D_{k,j+1}&=&(\cos\theta^{\prime}_{k-j-1}D_{k,j}+T\sin\theta^{\prime}_{k-j-1}e^{-i\varphi^{\prime}_{k-j-1}}B_{k,j}\,.\nonumber
\end{eqnarray}
The initial conditions are $ A_{k,0}=D_{k,0}=1$ and $ B_{k,0}=
C_{k,0}=0$.\\
As an example, we consider the generation of maximally entangled
four-photon state
\begin{eqnarray}
\Psi=\frac{1}{\sqrt{2}}(|0>_1|4>_2-|4>_1|0>_2)\,. \label{(23)}
\end{eqnarray}
We now give a simplified scheme. Consider the experiment, which is
shown schematically in Fig.2. A pair of photons from the
independent single photon source is incident on a symmetric beam
splitter. Behind the Beam Splitter, the state become an two-photon
path entangled state $ \Psi_1=\frac{1}{\sqrt{2}}(|0,2>-|2,0>)$. In
order to generate four-photon path entangled states, the spatial
separated photons are incident on two symmetric beam splitters
$BS_{2}$ and $BS_{3}$, respectively. The second input port of each
of these beam splitters is assumed to be single photon state
produced by single photon source. After the beam splitter $BS_{2}$
and $BS_{4}$ transformations, the two auxiliary photons are
measured and the outcome is accepted only when no photon was
detected by two photon detector $D_1$ and $D_2$. Thus, the state
is projected into $\Psi_2=\frac{1}{\sqrt{2}}(|1,3>-|3,1>)$. Then
this output state is taken as the input to the $BS_{4}$. Thus, we
can obtain the four-photon path entangled state
$|\Psi_3>=\frac{1}{\sqrt{2}}(|0,4>-|4,0>)$ with probability of
success $1/16$, which is slightly more than the probability of
success ($3/64$) for the scheme presented in Ref \cite{cer}.\\

In summary, we have suggested a feasible scheme to  generate
entangled state between two traveling modes for a fixed number of
photons, which is based on beam splitter transformation and
conditional zero-photon counters.  We also give a simplified
scheme to generate entangled four-photon quantum states with
probability of success $1/16$. This is slightly more than the
probability of success ($3/64$), which is achievable with the
scheme presented in Ref \cite{cer}. One of the difficulties of our
scheme in respect to an experimental demonstration is the
availability of photon number sources. Another difficulty consists
in the requirement on the sensitivity of the detectors. These
detectors should be capable of
distinguishing between no photon, one photon or more photons.\\

\newpage

\begin{flushleft}

{\Large \bf Figure Captions}

\vspace{\baselineskip}

{\bf Figure 1.} The schematic is shown to generate entangled state
between two traveling modes for fixed number of photon.
$BS_i(i=1,\cdots, N)$ denotes the beam splitters and $D_i$ are
photon number detectors. M denotes the mirror.

{\bf Figure 2.}This figure shows the simplified scheme of
entangled 4-photon quantum state generation. $BS_i$ denotes the
beam splitters and $D_i$ are photon number detectors. M denotes
the mirror.

\end{flushleft}
\end{document}